\DocumentMetadata{}
\documentclass[sigconf,10pt]{acmart}
\usepackage{amsmath,amsfonts}
\usepackage{algorithmic}
\usepackage{algorithm}
\usepackage{subcaption}
\usepackage{listings}
\usepackage{xspace}
\usepackage{color}
\usepackage{float}
\usepackage{xcolor}
\usepackage{textcomp}
\usepackage{stfloats}
\usepackage{url}
\usepackage{verbatim}

\setcopyright{rightsretained}

\copyrightyear{2024}
\acmYear{2024}
\setcopyright{rightsretained}
\acmConference[SoCC '24]{ACM Symposium on Cloud Computing}{November 20--22, 2024}{Redmond, WA, USA}
\acmBooktitle{ACM Symposium on Cloud Computing (SoCC '24), November 20--22, 2024, Redmond, WA, USA}
\acmDOI{10.1145/3698038.3698563}
\acmISBN{979-8-4007-1286-9/24/11}



\begin{document}

\title{ParaGAN: A Scalable Distributed Training Framework for Generative Adversarial Networks}

\author{Ziji Shi}
\orcid{0000-0001-9398-6507}
\affiliation{%
  \institution{National University of Singapore}
  \country{Singapore}
}
\email{zijishi@comp.nus.edu.sg}

\author{Jialin Li}
\orcid{0000-0003-3530-7662}
\affiliation{%
  \institution{National University of Singapore}
  \country{Singapore}
}
\email{lijl@comp.nus.edu.sg}

\author{Yang You}
\orcid{0000-0003-2816-4384}
\affiliation{%
  \institution{National University of Singapore}
  \country{Singapore}
}
\email{youy@comp.nus.edu.sg}

\begin{abstract}
Recent advances in Generative Artificial Intelligence have fueled numerous applications, particularly those involving Generative Adversarial Networks (GANs), which are essential for synthesizing realistic photos and videos. However, efficiently training GANs remains a critical challenge due to their computationally intensive and numerically unstable nature. Existing methods often require days or even weeks for training, posing significant resource and time constraints.

In this work, we introduce ParaGAN, a scalable distributed GAN training framework that leverages asynchronous training and an asymmetric optimization policy to accelerate GAN training. ParaGAN employs a congestion-aware data pipeline and hardware-aware layout transformation to enhance accelerator utilization, resulting in over 30\% improvements in throughput. With ParaGAN, we reduce the training time of BigGAN from 15 days to 14 hours while achieving 91\% scaling efficiency. Additionally, ParaGAN enables unprecedented high-resolution image generation using BigGAN.
\end{abstract}

\maketitle
\section{Introduction}\label{sec:intro}

The last decade has witnessed the success of Generative Adversarial Networks~\cite{goodfellow2014generative}, which has a wide range of applications including image super-resolution ~\cite{ledig2017photo}, image translation ~\cite{isola2017image, zhu2017unpaired}, photo inpainting ~\cite{yu2018generative, demir2018patch}. 
However, training GAN at scale remains challenging because of the computational demands and optimization difficulties.

Unlike other conventional neural networks where optimization is straightforward by taking gradient descents, there are two sub-networks to optimize in GAN, namely generator and discriminator. The generator samples from the noise and produces a fake sample as close to the real sample as possible, and the discriminator evaluates the generated sample. The generator aims to fool the discriminator, and the discriminator will try to identify the fake images from the real ones. Since the two components are optimized for two contradicting goals, it has been observed that GANs are difficult to converge. Therefore, to speed up the GAN training at large scale, we need a framework optimized on both system and numerical perspectives. 

\begin{table}
\small
  \caption{A summary of reported training time and model size for GANs trained on ImageNet dataset.}
  \label{tab:GAN}
  \begin{tabular}{cccc}
    \hline
    GANs & Hardware & Time & \# Params.\\
    \hline
      SNGAN~\cite{miyato2018spectral} & 8 V100 GPUs & 3d 13.6h & 81.44M\\
      ProgressiveGAN~\cite{karras2017progressive} & 8 V100 GPUs & 4d  & 43.2M \\
      ContraGAN~\cite{kang2020ContraGAN} & 8 V100 GPUs & 5d 3.5h  & 160.78M \\
      SAGAN~\cite{zhang2019selfattention} & 8 V100 GPUs & 10d 18.7h & 81.47M\\
      BigGAN~\cite{Brock2019LargeSynthesis} & 8 V100 GPUs & 15 d & 158.42M \\
    \hline
  \end{tabular}
\end{table}

\begin{figure}[t]
  \centering
  \includegraphics[width=\linewidth]{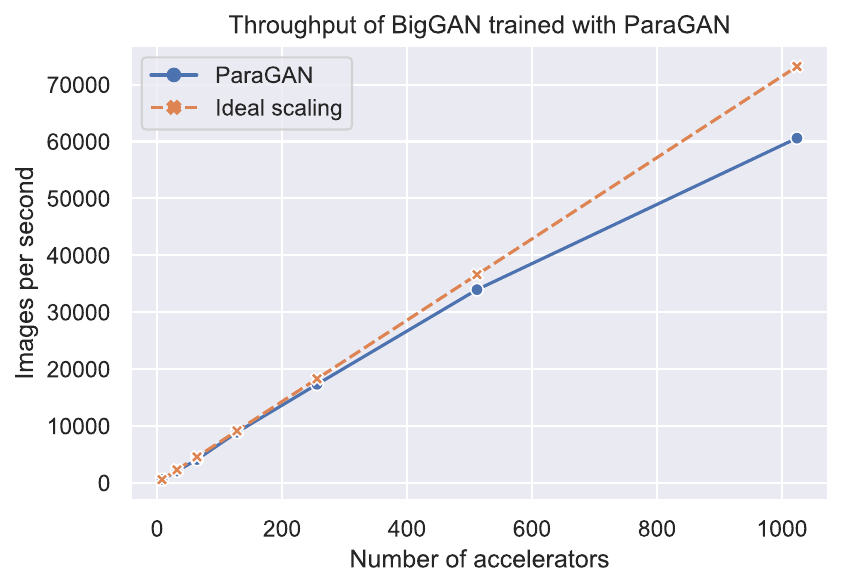}
  \caption{ParaGAN scales to 1024 TPU accelerators at 91\% scaling efficiency.}
  \label{fig:scaling}
\end{figure}

Due to the difficulty of optimizing GAN, many state-of-the-art GAN models take days or even weeks to train. For instance, BigGAN ~\cite{Brock2019LargeSynthesis} took 15 days on $8\times$ V100 GPUs to converge. Table \ref{tab:GAN} summarizes the reported training time of some of the state-of-the-art GAN models. This makes it difficult to quickly reproduce, evaluate, and iterate GAN experiments. Also, current GAN frameworks usually only support training on a small number of nodes~\cite{kang2020ContraGAN, 2021mmgeneration, lee2020mimicry}. 

We argue that training speed is an important yet often ignored factor in the current GAN training landscape, and we propose to accelerate it with distributed training. However, distributed GAN training has several challenges. First of all, most data centers have storage nodes and compute nodes separated for elasticity, but network congestion can happen from time to time, which prolongs the latency between nodes and affects training throughput.  
Secondly, there are usually different types of accelerators in the data center, but each of them has its architectural design and preferred data layout. If ignored, it could lead to under-utilization of accelerators. 
Last but not least, training GAN at scale may cause a convergence problem, in which the GAN loss does not converge to a stable equilibrium. Therefore, this framework has to consider both system and numerical perspectives. 

In this work, we present \textbf{ParaGAN}, the first distributed training framework that supports large-scale distributed training for high-resolution GAN. We identify the performance bottlenecks when training at scale and optimize them for efficiency. ParaGAN has a simple interface for building new GAN architecture, and it supports CPU, GPU, and TPU. 

The main contributions of ParaGAN include: 
\begin{itemize}
    \item We design and implement the first scalable distributed training framework for GAN with optimizations on both system and numerical perspectives. With ParaGAN, the training time of BigGAN can be shortened from 15 days to 14 hours with 1024 TPU accelerators at 91\% scaling efficiency, as shown in Fig.~\ref{fig:scaling}. ParaGAN also enables direct photo-realistic image generation at unprecedented $1024\times1024$ resolution, which is $4\times$ higher than the original BigGAN model.
    
    \item From the system optimization perspective, we use congestion-aware data pipeline and hardware-aware layout transformation to improve the accelerator utilization, and low-precision training to alleviate the memory stress. They contribute to 30-40\% throughput improvements over baseline. 
    
    \item From the numerical optimization perspective, we show that the generator and discriminator can be optimized independently, and present an asynchronous update scheme together with an asymmetric optimization policy. 
\end{itemize}

The paper is organized in the following manner: we discuss the motivation and requirement for large-scale GAN training in Section~\ref{sec:motivation}; in Section~\ref{sec:design}, we will explain our design for ParaGAN, and how those architectural considerations can address the requirements; in Section~\ref{sec:imp} and Section~\ref{sec:algo_opt}, we will cover the system-level and numerical-level optimizations for scalable training in ParaGAN respectively; in Section~\ref{sec:eval}, we present our scalability evaluation of ParaGAN and study the effect of different optimization techniques; a brief review of related work on GAN and large-scale distributed training is presented in Section~\ref{sec:related_work}; we conclude this study of GAN in Section~\ref{sec:conclude}.
\section{Motivation and Requirements}\label{sec:motivation}
\begin{figure}[t]
  \centering
  \includegraphics[width=\linewidth]{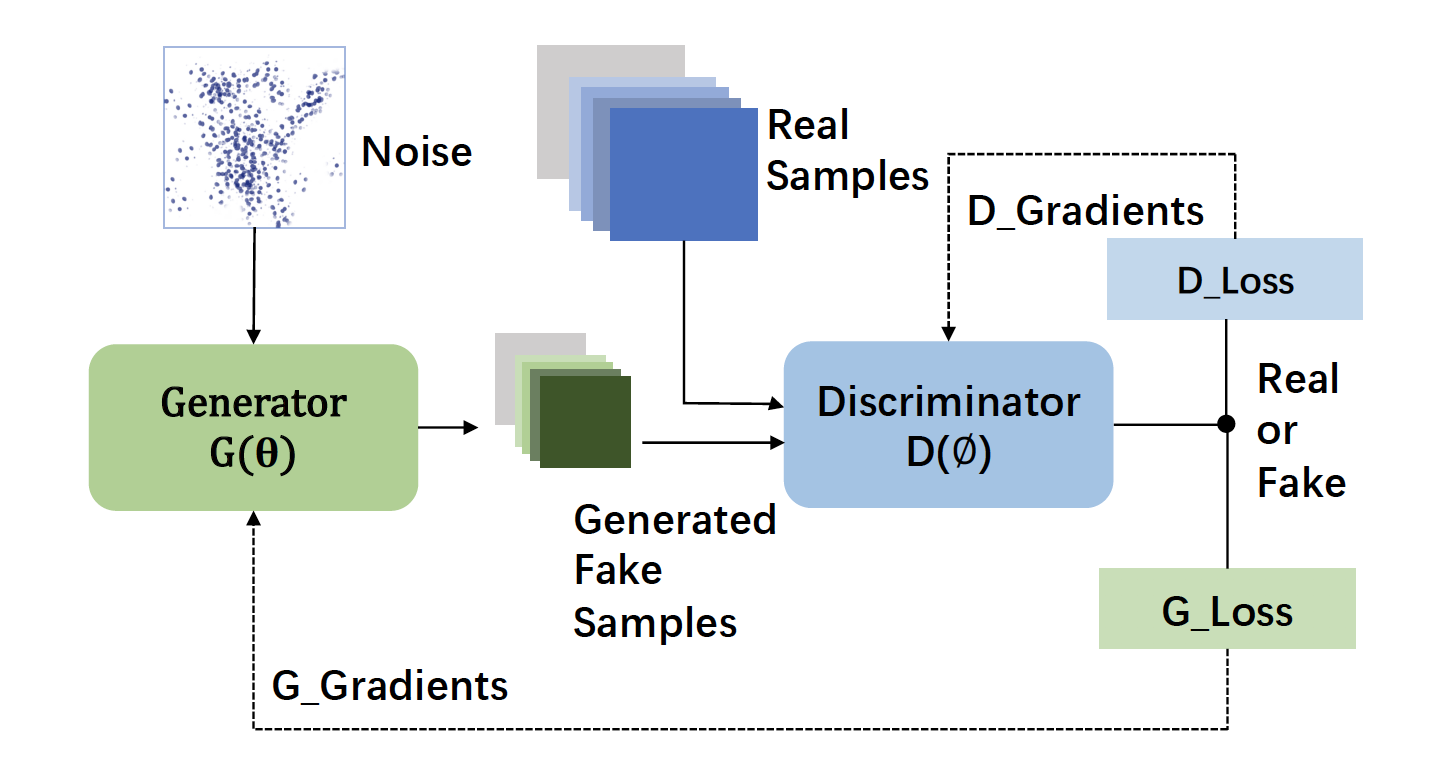}
  \caption{Typical GAN architecture.}
  \label{fig:gan}
\end{figure}

We begin by considering the basic components of GAN to discover the key requirements for GAN training. As shown in Fig. \ref{fig:gan}, a GAN consists of a generator and a discriminator. The generator generates fake data samples, while the discriminator distinguishes between the generated samples and real samples as accurately as possible. The learning problem of GANs is a minimax optimization problem. The goal of the optimization is to reach an equilibrium for a two-player problem:

\[\min_{G}\max_{D}\mathbb{E}_{x \sim q_{data}(x)}\left[\log{D(x)}\right]+\mathbb{E}_{z \sim p(z)}\left[\log{(1-D(G(z)))}\right]
\]
where ${z\in \mathbb{R}^{d_z}}$ is a latent variable drawn from distribution $p(z)$. The discriminator seeks to maximize the sum of the log probability of correctly predicting the real and fake samples, while the generator tries to minimize it instead. The convergence of GAN is defined in the form of Nash Equilibrium: one network does not change its loss regardless of what the other network does.

Since the two networks have contradicting goals, the training process of GAN is a zero-sum game and can be very unstable. 
Recent works show that: i) GAN may converge to points that are not local minimax using gradient descent, in particular for a non-convex game which is common ~\cite{jin2020local,daskalakis2018limit}, and ii) gradient descent on GAN exhibits strong rotation around fixed points, which requires using very small learning rates ~\cite{Mescheder2017TheGANs, balduzzi2018mechanics}. Also, GANs training is sensitive to the hyperparameters and initialization ~\cite{Lucic2018AreStudy}. Therefore, it is observed that GANs are difficult to optimize, and this is also the reason why it takes a long time to train them.

There are some existing GAN libraries ~\cite{kang2020ContraGAN, 2021mmgeneration, Lucic2018AreStudy, lee2020mimicry} for training state-of-the-art GANs. They provide standardized building blocks like network backbone and evaluation metrics, making it easy to build new models. However, they focus less on the system performance, and training GAN can still take days if not weeks. If the training process can be massively paralleled, the iteration cycles can be greatly shortened.
Motivated by the challenges with the status quo, we outline the following requirements for a desirable distributed GAN training framework:

\textit{Performance and Scalability}. Performance is of critical importance to GAN training. The experiments from ~\cite{Lucic2018AreStudy} in total took 60K GPU hours to train, making it hard to reproduce and improve over. 
Completing such a workload is almost infeasible without using distributed training. 
Also, since most GAN models are compute-intensive and can be fitted into the memory of one worker, data parallelism can be used as the distributed training strategy, by placing identical copies of the model on each accelerator worker. The framework should be scalable as the number of accelerators scales.

\textit{Numerical Stability}. The convergence of GAN can be very volatile. The framework should have means to stabilize the training process. Also, the framework should support large batch training as the number of accelerator scales.

In ParaGAN, our approach is to co-design the solution on the system and optimization levels. At the system level, we pinpoint performance bottlenecks, primarily arising from network congestion and sub-optimal accelerator utilization. To alleviate these issues, we introduce a congestion-aware data pipeline and implement a hardware-aware layout transformation. On the optimization front, we argue that the training process of the discriminator and generator can be treated independently, and we find that decoupling the training of the generator and discriminator offers advantages. To capitalize on this insight, ParaGAN employs an asynchronous update scheme along with an asymmetric optimization policy.

\section{Design and Architecture}\label{sec:design}

In this section, we will give an overview and discuss the design decisions of ParaGAN. We recognize that the scalability is usually limited by the latency between nodes. Furthermore, when scaling up the GAN training, the numerical instability problem happens more often. We therefore divide the discussions into two folds and present our co-designed approach for system throughput and training stability as it scales.

\subsection{Programming Model}

\begin{figure}[t]
\includegraphics[width=\linewidth]{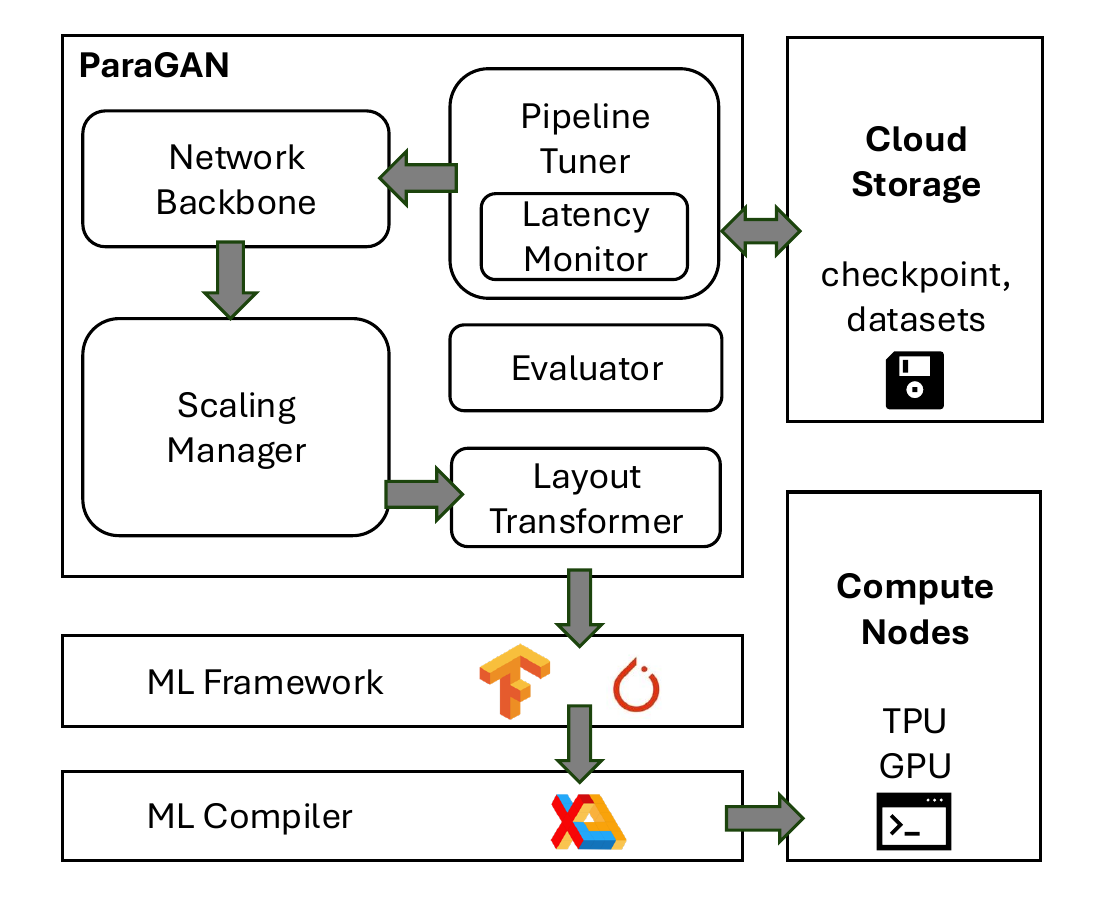}
\centering

\caption{Overview of ParaGAN.}
\label{fig:overview}
\end{figure}

The architecture of ParaGAN is presented in Fig.~\ref{fig:overview}. ParaGAN is implemented on top of TensorFlow, but it can be ported to support other DL frameworks. ParaGAN provides high-level APIs for GAN which include scaling manager, evaluation metrics, and common network backbones. Users of ParaGAN can import from ParaGAN or define their components. ParaGAN then performs layout transformation on tensors and invokes TensorFlow, which converts the model definition to a computational graph. An optional XLA~\cite{50530} pass can be performed followed by that. After that, the training loop starts on the compute nodes that host accelerators like GPUs and TPUs.

\begin{lstlisting}[language=Python, caption={Parallel GAN Training and Evaluation}, label={code:pg}, float=t]
import paragan as pg

class Generator:
    def model_fn(x, y):
    # generator model
    return output

class Discriminator:
    def model_fn(x, y):
    # discriminator model
    return output, out_logits

scaling_mgr = pg.ScalingManager(config=cfg, batch_size=2048, num_workers=128)

g = Generator()
d = Discriminator()
model = pg.Estimator(g,d)

# training
for step in cfg.max_train_steps:
     scaling_mgr.train(gan)

# evaluation
scaling_mgr.eval(metric="fid")
\end{lstlisting}\label{code:pg}

We introduce three concepts in ParaGAN: 

\subsubsection{Scaling Manager}

The scaling manager is in charge of hyper-parameters that need to be tuned when scaling, including learning rate, optimizer, and local batch size. Users can use the best hyper-parameters from a single worker as a starting point, and ParaGAN will scale them based on the number of workers and learning rate schedules. Users can also define their scaling manager.

\subsubsection{Network Backbones}
Users usually start by building upon existing GAN architectures. We also provide some popular GAN architectures as backbone, including but not limited to:

\begin{itemize}
    \item BigGAN ~\cite{Brock2019LargeSynthesis};
    \item Deep Convolutional GAN (DCGAN) ~\cite{radford2015unsupervised};
    \item Spectral Norm GAN (SNGAN) ~\cite{miyato2018spectral}
\end{itemize}

\subsubsection{Evaluation Metrics}
Evaluation metrics can be implemented differently across papers, and this can cause inconsistency. We provide commonly used evaluation metrics including Frechet Inception Distance (FID) and Inception Score (IS).

\subsection{Computation Model}

Training on the cloud usually involves host machines (mostly CPU nodes), compute nodes and storage nodes.
As depicted in Fig. \ref{fig:overview}, the host fetches input data from the storage node, builds the model, chooses the tensor layout for the target accelerator, and feeds to the compute nodes. 
After that, the accelerators execute forward and backward passes and then synchronize gradients among other accelerators. 
At the same time, the host prefetches and transforms the data from the storage node. 
Model checkpoints are saved to the storage node which can be an object store or mounted file system. 

A host machine can be a standalone node, while most of the time it co-locates with the compute nodes in the same chassis, and is connected to the accelerators via high-speed PCI-e bus or NVLink. Meanwhile, the host-to-host and host-to-storage connections usually go through Ethernet. 

\subsection{System Optimizations}
To satisfy the scalability requirement, we design ParaGAN with optimizations on data pipeline, computation, and memory.

We optimize the data pipeline performance by using a congestion-aware data pipeline. For data centers, the compute and storage nodes are usually physically distributed and interconnected via Ethernet instead of high-speed InfiniBand. The network traffic between them may not always be stable since the infrastructure is shared with other tenants. This issue is further pronounced when the number of participating data parallel workers scales up because data parallelism is a synchronous training method that is sensitive to the latency of the slowest participant. Therefore, ParaGAN continuously monitors the data pipeline latency and implements a congestion-aware data pipeline tuner.

To achieve a higher accelerator utilization, ParaGAN performs hardware-aware layout transformation. A data center usually has multiple types of accelerators, and different accelerators have different micro-architectures and instructions, thereby having different preferred data layouts. For example, Nvidia A100 GPUs prefer half-precision data in multiples of 64, and single-precision data in multiples of 32, while previous generations prefer multiples of 8. For TPU, the preferred data layout should have a multiple of $128$ on the lane dimension and $8$ on the sublane dimension. If misconfigured, this will result in unnecessary padding which reduces accelerator utilization and increases memory consumption. We come up with the hardware-aware layout transformation to transform the data into an accelerator-friendly format to maximize accelerator utilization.

Memory usage can be reduced by mixed-precision training with Brain Float 16 (bf16) format. Theoretically, using purely bf16 can save half of the memory for activation. However, we observe through experiments that bf16 is not suitable for the layers that are sensitive to overflow/underflow. To be precise, we found that the generator and discriminator's last layer are more sensitive to precision. As such, we apply full precision (fp32) on these layers.

\subsection{Numerical Optimizations}

Another key contribution of ParaGAN is using asymmetric training to stabilize GAN. As the number of workers scales, a larger batch size can accelerate training. However, we observe that the performance of large batch training for GAN is not as stable, and mode collapse happens frequently. Since mode collapse is a type of GAN failure raised due to a highly coupled optimization process, to solve this problem, ParaGAN comes with an asymmetric optimization policy and asynchronous update scheme to decouple the process. 
 \section{Implementation}\label{sec:imp}

\begin{figure}[t]
  \centering
  \includegraphics[width=\linewidth]{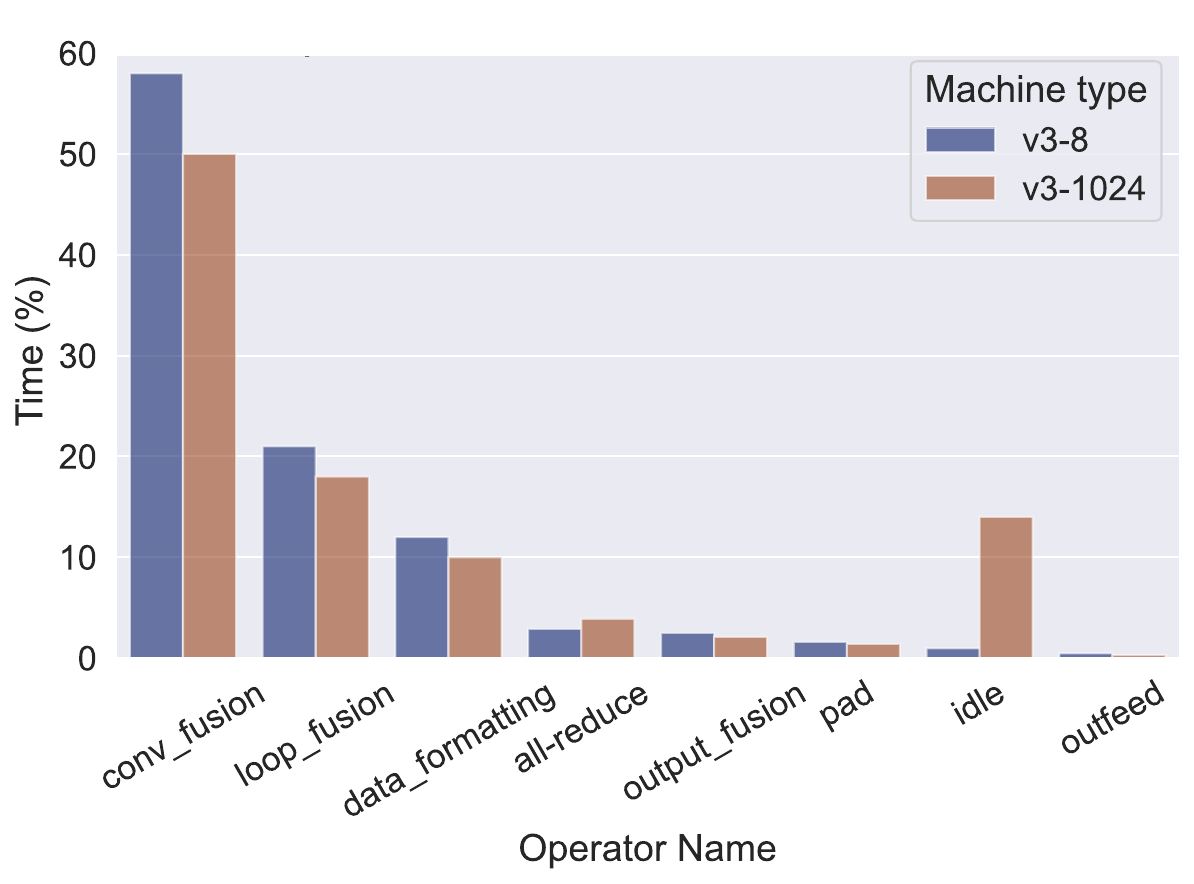}
  \caption{Operator usage profile when training at scale.}
  \label{fig:profile}
\end{figure}

To begin with, we profile BigGAN training on native TensorFlow ~\cite{Lucic2018AreStudy} and present the result in Fig. \ref{fig:profile}. As we increase the cluster size from 8 to 1024 TPU workers, idle time significantly increases due to increased communication, but convolution operation still makes up most of the time. It indicates that training GAN is a compute-bound workload. Therefore, we focus on improving the accelerator utilization in ParaGAN.

To achieve this goal, we use congestion-aware data pipelining to reduce data pipeline latency, hardware-aware layout transformation to increase accelerator utilization, and mixed-precision training with bfloat16 for reduced memory.

\subsection{Congestion-Aware Data Pipelining}

The communication between the compute node and storage node is much slower compared to the on-chip interconnect. Different from TPU-TPU communication using high-speed onboard interconnect or GPU-GPU communication that could go through NVLink/PCI-e bus, communication between the accelerator node and cloud storage goes through Ethernet, which is usually orders of magnitude slower than the former. 

Furthermore, during GAN training, a huge amount of data will be transmitted via Ethernet. For instance, the ImageNet 2012 dataset is over 150 GB, and training BigGAN to convergence takes around 240 epochs, resulting in a total of 32.74 TB of data being transmitted during the training phase. When the number of workers increases, the amount of peak data transmitted increases at the same rate, since data parallel training will send the same amount of data to each worker. This further exacerbates the congestion problem. 

Last but not least, network jittering can significantly affect the training throughput. We observe that due to traffic congestion within the data center, the latency between the storage node and the accelerator node is not always stable during peak hours, and it can result in low accelerator utilization when the data pipeline cannot provide enough samples to saturate the accelerator's compute capability.

Although both TensorFlow and PyTorch implement data pipelines to hide the data loading latency, when severe network jittering happens, data loading and pre-processing take much longer than usual, and it can be a bottleneck in large-scale distributed training. As shown in Fig. \ref{fig:profile}, when the number of workers scales from 8 to 1024, it spends 13.6\% more time on idling, while data outfeeding time stays close. This indicates that the accelerators are busy waiting for data infeed and gradient synchronization, which leads to reduced utilization. 

ParaGAN dynamically adjusts the number of processes and size of the pre-processing buffer in response to the high-variance network. It is implemented by maintaining a sliding window for network latency during runtime. If the current latency over the window exceeds the threshold, ParaGAN will increase the number of threads and buffer for pre-fetching and pre-processing; once the latency falls below the threshold, it releases the resources for pre-processing. This may come at the expense of increased shared memory usage, but shared memory is usually abundant during model training. 

Saving model checkpoints will also go through the host. We use an asynchronous checkpoint writer to save model checkpoints. The checkpoint will be streamed into the output buffer instead of having a blocking call to pass it to the CPU host.

\subsection{Hardware-Aware Layout Transformation}

Zero-padding is frequently used in GAN when the input cannot fit into the specified convolution dimension. For example, a matrix of shape [100, 100] will need 6384 zeros padded to run on a $128\times128$ matrix unit, which wastes 39\% computing resources. As such, zero-padding hinders the accelerator performance because memory is wasted by padding, leading to lower accelerator and memory utilization rates.

Given that there exists an accelerator-dependent format requirement, ParaGAN performs batching opportunistically on the input data. In NCHW (batch size x number of channels x height x width) format, ParaGAN tries to batch them such that N/H/W are multiple of 128 before running on TPU so that the accelerator memory can be efficiently utilized. 

On top of the batch dimensions, ParaGAN also seeks opportunities to batch intermediate results to be a multiple of optimal layout dimensions. Such opportunities can be found at \textit{reshape} and \textit{matmul} operators. For instance, if two input matrices are to multiply the same weight, we can concatenate the two input matrices before the matrix multiplication operation to save kernel launch overhead.

\subsection{Mixed-Precision Training}

ParaGAN supports mixed-precision training with Bfloat16~\cite{bf16，kalamkar2019study} (bf16) format, which has a lower memory footprint compared to double precision format, allowing users to use a larger batch size or fit a bigger model into memory.
However, porting bf16 while maintaining convergence is not straightforward because bf16 trades floating-point precision for range. As a result, hyperparameters with smaller values will also need to be adjusted to accommodate lower-precision bits. For example, $\epsilon$ in Adam optimizer is a small value added to the denominator to avoid zero-division. With low-precision bits, it is necessary to use a slightly larger $\epsilon$ value.

Throughout our evaluations, we observe that weights and gradients are more sensitive to the bf16 format, while activation can be represented using lower precision without significantly affecting the coverage. Also, the shallow layers are less sensitive compared to deeper layers. We provide bf16 as an option for users to explore.

\section{Numerical Optimizations}\label{sec:algo_opt}

Synchronous large-batch training is a key technique for accelerating neural network training. However, a major issue with using large batches is that they can lead to numerical instability. In the context of GAN, numerical instability can cause mode collapse or divergence, failing to generate realistic samples with good variety. 

A key reason for the numerical instability is that GAN is a two-player game. As such, when the training proceeds, the generator and discriminator can get tightly optimized towards each other. Meanwhile, the generator and discriminator are two networks with different learning dynamics. We argue that the optimization process for GAN should treat them differently.

Therefore, we propose an asynchronous update scheme to decouple the generator and discriminator, and an asymmetric optimization policy for optimizing the two differently.

\subsection{Asynchronous Update Scheme}
The optimization of GAN is traditionally a serial process where the generator (G) and discriminator (D) update one after another. We question the necessity of an iterative process and propose an asynchronous update scheme. 

\begin{figure}[t]
  \centering
  \includegraphics[width=\linewidth]{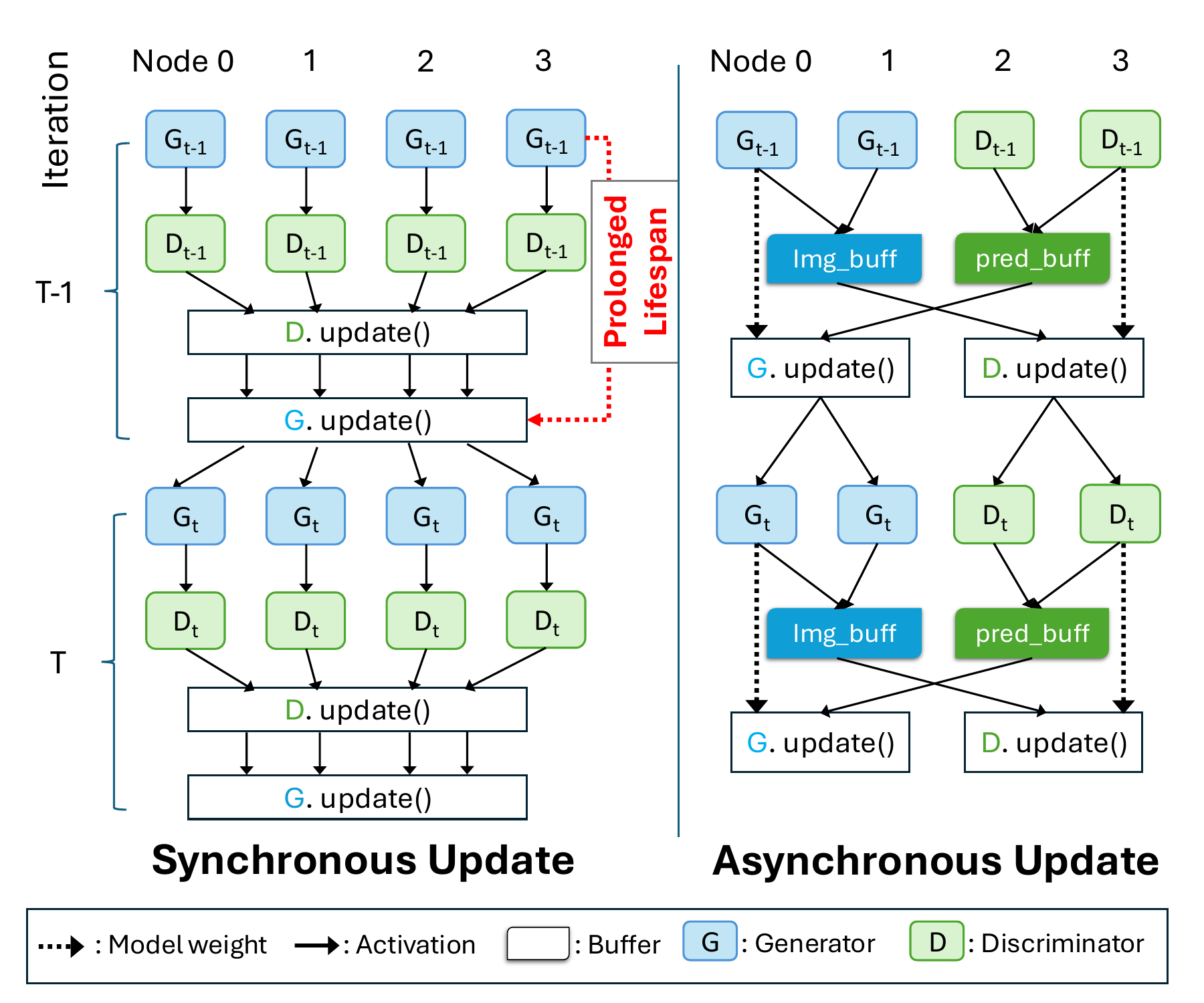}
  \caption{Synchronous update and asynchronous update scheme in ParaGAN. $G_t$ and $D_t$ are the model weights of the generator and discriminator at iteration t. }
  \label{fig:async_gan}
\end{figure}

As shown in Fig.~\ref{fig:async_gan}, the discriminator will not update its parameter until the generator finishes backpropagation, and vice versa. This is because of the data dependency: D reads both the real sample and the generated sample as input to calculate D loss, while G loss depends on the output and weight of D. The data dependency issue requires both subnetworks to only update one after another, which leads to prolonged tensor lifespan due to blocking, and frequent context switching when updating one or another.

Through our experiments, we find that the discriminator is less sensitive to changes in the generator output. In other words, the discriminator can still perform well even if its input comes from the generator of the previous iteration. This is because the primary goal of the discriminator is to distinguish the generated sample from real samples, and it is an easier task compared to that of the generator. This has motivated us to update the discriminator asynchronously. 

As shown in Fig.~\ref{fig:async_gan} (right), ParaGAN proposes an asynchronous update scheme: instead of waiting on the other component, the generator/discriminator can write their intermediate output to the buffer and proceed to update using the \textit{current} state of the network. For iteration $t$, discriminators $D_t$ receive a batch of real and generated samples from the image buffer (\texttt{img\_buff}). Similarly, the generators can use the snapshot of the current discriminator state and D predictions in \texttt{pred\_buff} to calculate to gradient for backpropagation, breaking the data dependency. It is therefore possible to run both generator and discriminator in parallel on different nodes. Furthermore, the generator and discriminator ratio is now adjustable thanks to the decoupled design, making it possible to apply different batch sizes for both parties.

\subsection{Asymmetric Optimization Policy}
Generators and discriminators have different neural network architectures and play opposed roles during training.
Thus, they are two entities with significantly distinct numerical properties.
It is observed that the discriminator is more stable than the generator.
This indicates that the generator and discriminator should be treated differently by using different sets of optimization techniques.
However, previous researchers treat them as the same in terms of numerical optimization.
We thus propose an Asymmetric Optimization strategy for generator and discriminator in GAN training.

To accomplish this goal, ParaGAN firstly implements some of the latest work on optimizers including Adabelief~\cite{Zhuang2020AdaBeliefGradients}, rectified Adam (RAdam) ~\cite{liu2020variance}, Lookahead ~\cite{Zhang2019LookaheadBack}, and LARS ~\cite{you2017large}. We then empirically explored different optimizers for training GAN, and we found that there may not be a single clear winner for all GAN architectures. 

\begin{figure}[t]
\includegraphics[width=\linewidth]{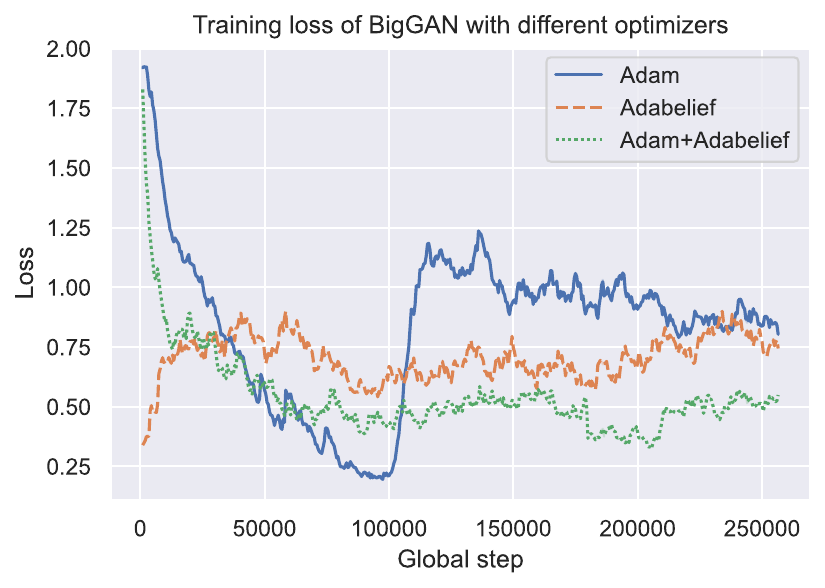}
\centering
\caption{Effect of different optimizer policies. Adam+Adabelief: discriminator trained using Adam and generator trained using Adabelief.}
\label{fig:opm_loss}

\end{figure}

However, through experiments, we observe that it can be beneficial to use different optimizers for the generator and discriminator respectively. Fig.~\ref{fig:opm_loss} shows that although Adam reaches the lowest loss within 100K steps, it collapses thereafter, which is not desirable as it indicates the training has not reached the stable equilibrium. Adabelief is a more adaptive variant of Adam optimizer, and it can adjust the size of the weight update based on a comparison with previous updates. Our experiments also validate that Adabelief outperforms Adam. However, when using an asymmetric pair of optimizers (Adabelief for the generator and Adam for the discriminator), the training process can converge to a better equilibrium point (lower loss), and the training process is more stable (flatter loss curve), especially towards the end of the training. 

We conjecture that the reason is that the two networks have different learning dynamics, thus they require different optimizers for gradient descent. The generator aims to learn from the real-world distribution, thus it must provide a good variety, and the optimizer for the generator should be more agile. However, the task for the discriminator is relatively easy: predicting the label (\textit{real} or \textit{fake}) for the supplied images. Therefore, it is required to make consistently good predictions and be robust to changes. As such, we believe using the asymmetric optimizer policy can combine the best of both worlds.

In ParaGAN, users can set the optimization policy for the generator and discriminator respectively, which currently includes optimizers, learning rate schedulers, warmup epochs, and gradient norms. 

\section{Evaluation}\label{sec:eval}
In this section, we aim to answer the following questions: 1) how is the performance of ParaGAN compared to other frameworks? 2) how much does each part of the system contribute to the overall performance? And 3) how do the numerical optimizations improve convergence?

In this section, we first evaluate the end-to-end performance of ParaGAN using three metrics:

\begin{itemize}
    \item \textbf{steps per second} measures the number of steps ParaGAN can train per second;
    \item \textbf{images per second} measures the throughput of ParaGAN trained with the ImageNet 2012 dataset;
    \item \textbf{time to solution} measures the time it takes to reach 150k steps on ImageNet at $128\times128$ resolution.
\end{itemize}

We first compare ParaGAN with other popular frameworks for end-to-end performance (Sec.~\ref{sec:framework}) and evaluate the scaling efficiency for ParaGAN (Sec.~\ref{sec:scaling_exp}). We ablate the optimizations performed on ParaGAN in Sec~\ref{sec:abs_study}.

\subsection{Experiment Setup}
We choose the BigGAN model and train it on the ImageNet ILSVRC 2012 dataset as our evaluation method because of BigGAN's profound impact on high-resolution image generation and its high computational requirements (Table \ref{tab:GAN}), and ImageNet's wide variety of classes (1000 classes) also presents a significant training challenge. The evaluations were conducted using V100 GPU and TPU v3. 

While we use BigGAN to benchmark ParaGAN, our framework is generally applicable to other GAN architectures and datasets, and it is not tightly coupled with any specific accelerator backends.

\subsection{Framework-level Experiments}\label{sec:framework}
\begin{figure}[t]
\includegraphics[width=\linewidth]{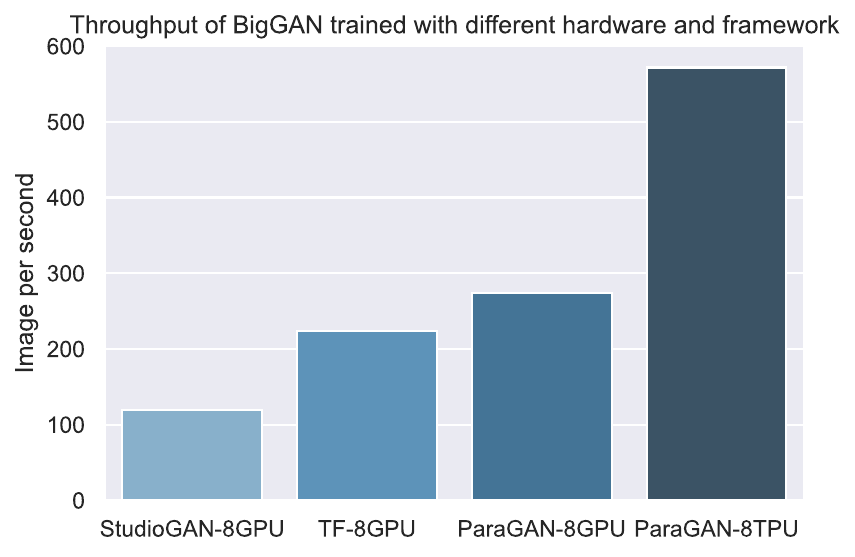}
\centering
\caption{Throughput of different systems and hardware combinations.}
\label{fig:gpu_tpu}
\end{figure}

We compare ParaGAN with StudioGAN ~\cite{kang2020ContraGAN} and native TensorFlow ~\cite{Lucic2018AreStudy} for GPU performance in Fig. \ref{fig:gpu_tpu}. For each of the experiments, BigGAN is trained on ImageNet $128\times128$ resolution. We use $8\times$ Tesla V100 GPUs except for ParaGAN-8TPU setting which employs 8 TPU. 

We observe that ParaGAN outperforms both the native TensorFlow and StudioGAN with 8 GPUs. We conjecture that the performance gain on the GPU setting is mainly attributed to the use of congestion-aware data pipeline and hardware-aware layout transformations. We also observe that the performance gap is further pronounced when switching to the TPU as the accelerator. The following evaluations use the TPU as the accelerator unless otherwise specified.

\subsection{Scaling Experiments}\label{sec:scaling_exp}
We will discuss the strong and weak scaling results in this section. In the strong scaling experiments, we keep the total workload constant and vary the number of workers to examine the speedup on time-to-solution. Whereas in the weak scaling experiments, we keep the per worker workload (\textit{batch size per worker}) constant and increase the number of workers. 

\subsubsection{Strong Scaling}
For strong scaling experiments, we fix the total batch size to be 512 and train for 150k steps as target workload. Note that to be consistent with other experiments, we train on BigGAN at $128\times128$ resolution, which is smaller than the model trained in Fig.~\ref{fig:scaling}. We aim to study the effect of decreased per-worker workload when scaling. 

\begin{figure}[t]
    \centering
    \begin{subfigure}[b]{\linewidth}
        \centering
        \includegraphics[width=0.8\linewidth]{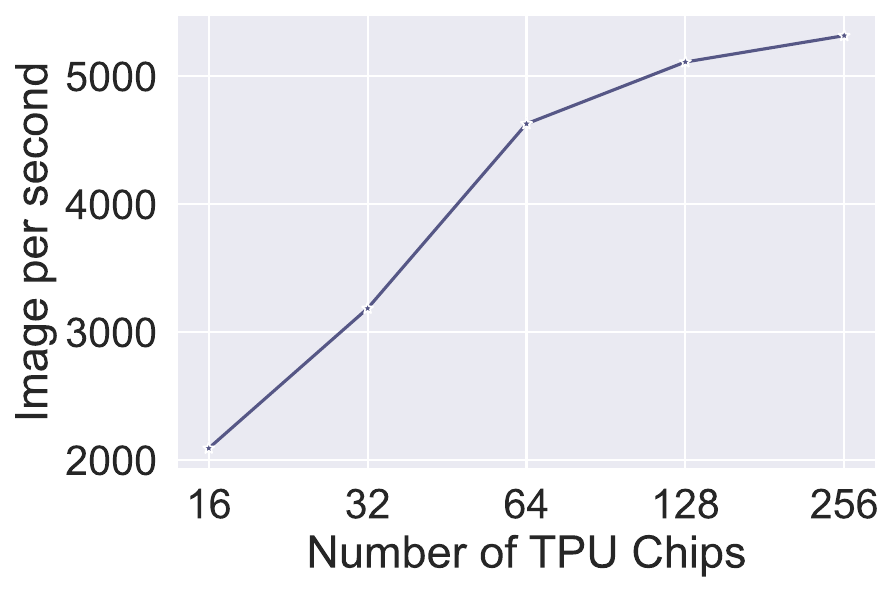}
        \caption{Image per second.}
        \label{fig:strong_thpt}
    \end{subfigure}
    \hfill
    \begin{subfigure}[b]{\linewidth}
        \centering
        \includegraphics[width=0.8\linewidth]{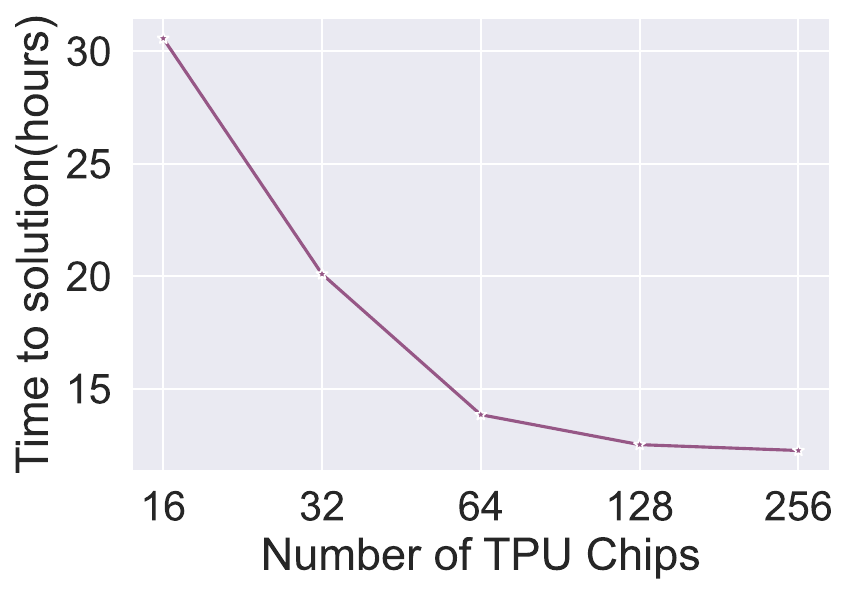}
        \caption{Time-to-solution.}
        \label{fig:strong_tts}
    \end{subfigure}
    \caption{Strong scaling with ParaGAN. Each TPU chip has two accelerators.}
    \label{fig:strong_scaling}
\end{figure}

As can be seen from Fig~\ref{fig:strong_scaling}, with an increasing number of workers, the time to solution decreases from over 30 hours to 3 hours. We note that the scaling efficiency drops from 128 to 512 workers (64 to 256 TPU chips). This is because as we fix the global batch size to 512, the per-worker workload drops from 4 samples to 1 sample per batch, which under-utilizes the TPU. Thus, the time spent on communication overweights the computation when the batch size is too small. This is also verified by Fig~\ref{fig:strong_scaling}, where the image per second barely improves with an increasing number of accelerator workers. However, when the workload can saturate the accelerator, the scaling efficiency can be near optimal as shown in Fig.~\ref{fig:scaling}.

\begin{figure}
     \centering
     \begin{subfigure}[b]{\linewidth}
         \centering
         \includegraphics[width=0.8\linewidth]{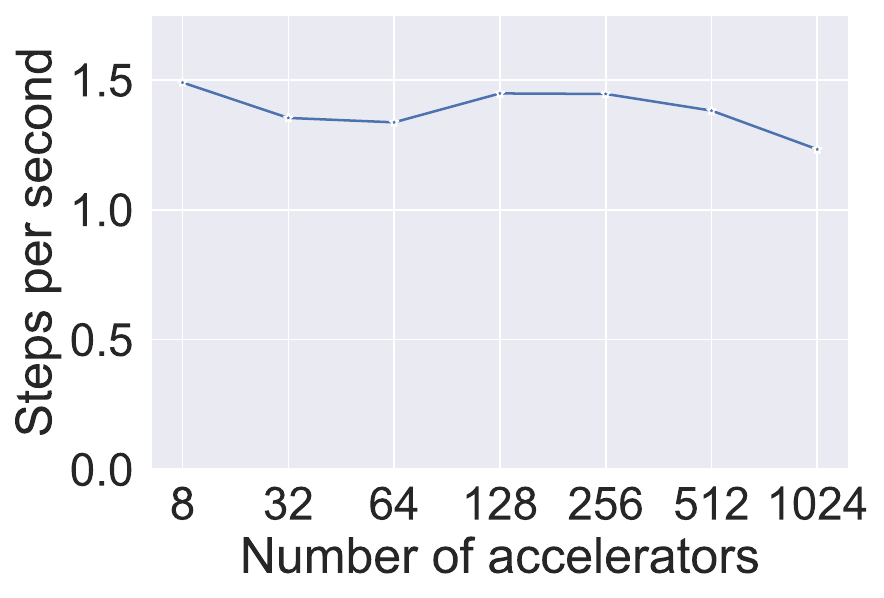}
         \caption{Step per second.}
         \label{fig:weak_step}
     \end{subfigure}
     \hfill
     \begin{subfigure}[b]{\linewidth}
         \centering
         \includegraphics[width=0.8\linewidth]{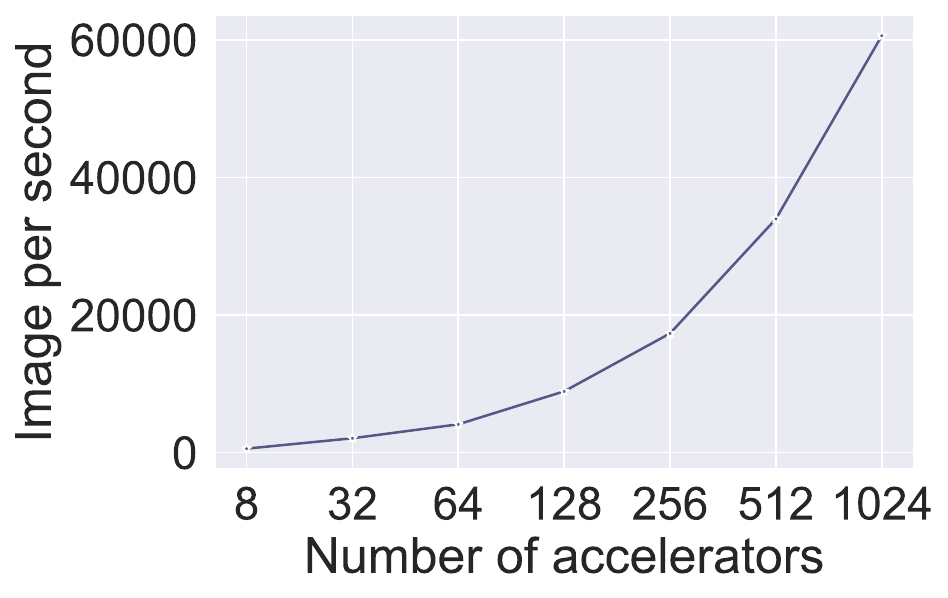}
         \caption{Image per second.}
         \label{fig:weak_ips}
     \end{subfigure}
    \caption{Weak scaling with ParaGAN.}
    \label{fig:weak_scaling}
\end{figure}

\subsubsection{Weak Scaling}
In the weak scaling experiments, we fix the batch size per worker and evaluate the performance of our framework by increasing the number of workers. Firstly, we find the largest batch size for a single accelerator that does not lead to out-of-memory error. Then, we scale the total batch size proportionally concerning the number of workers. Therefore, the amount of workload per worker is kept identical. The weak scaling experiments examine how well ParaGAN can handle communication with an increasing number of workers. 

As can be seen in Fig.~\ref{fig:weak_scaling}, the trend in step-per-second is relatively steady even when using 1024 workers. It shows that ParaGAN can scale out well to a large number of workers while keeping a high scaling efficiency. It is worth noting that, as the number of workers scales, the system will be more likely to suffer from network jittering and congestion. A relatively flat curve (Fig.~\ref{fig:weak_step}) indicates that the data pipeline optimization in ParaGAN is effective in case of congestion.

\subsection{Accelerator Utilization}\label{sec:hw_util}

The basic computing unit of TPU is MXU (matrix-multiply unit). The utilization of MXU measures the time MXU is being occupied, and higher utilization is more desirable. 

\begin{figure}[t]
\includegraphics[width=\linewidth]{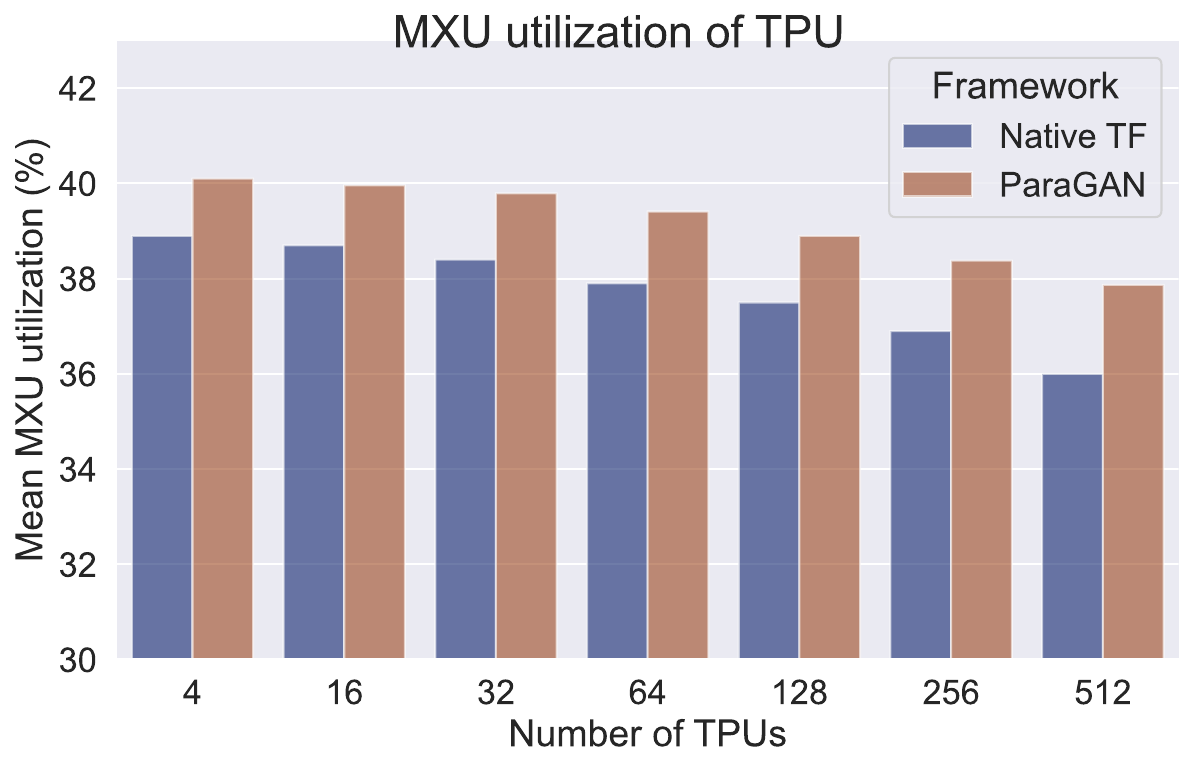}
\centering
\caption{Accelerator utilization of BigGAN trained with native TensorFlow and ParaGAN.}
\label{fig:mxu_utils}
\end{figure}

We compare the accelerator utilization of BigGAN 128x128 on baseline ~\cite{Lucic2018AreStudy} and ParaGAN. Fig. \ref{fig:mxu_utils} shows that ParaGAN outperforms native implementation with higher MXU utilization across different TPU configurations. We wish to highlight that even 2\% improvement can be important when scaling to thousands of workers. 

It is also worth noting that, with an increasing number of accelerators, the amount of communication increases, but ParaGAN can maintain a relatively higher utilization than native implementation, and the gap is increasing. It indicates that computation still dominates the training time as compared to native TensorFlow, and ParaGAN can keep up with scaling out.

\subsection{Ablation Study}\label{sec:abs_study}

We present the ablation study for the system optimizations in Table \ref{tab:abl_study}. For the numerical optimizations, since the convergence also depends on the underlying GAN architecture and datasets, we leave them to be decided by the users. Results in Table \ref{tab:abl_study} are collected on BigGAN trained on ImageNet at 128x128 resolution on 128 TPUv3 accelerators, under the same batch size (2048) from the original paper ~\cite{Brock2019LargeSynthesis}.

\begin{table*}[t]
  \caption{Ablation Study of System Optimizations.}
  \label{tab:abl_study}
  \begin{tabular}{c|cccc|c}
    \hline
    Effective & \multicolumn{4}{c|}{System Optimizations} & Img/sec \\
    Batch Size & Data Pipelining & Layout Transformation & Mixed-Precision &\\ 
    \hline
    2048 & & & & & 6459 \\
    2048 & \checkmark & & & & 7158 (\textbf{+10.8\%}) \\
    2048 & \checkmark & \checkmark & & & 7412 (\textbf{+3.9\%}) \\
    2048 & \checkmark & \checkmark & \checkmark & & 8539 (\textbf{+15.2\%}) \\
    \hline
  \end{tabular}
\end{table*}

\textbf{Data pipeline} provides 8-15\% performance improvement over the baseline. When the number of accelerators increases, network jitter caused by congestion is more likely to happen, making data loading the weakest point in the training process. In ParaGAN, we try to saturate the accelerators by dynamically adjusting the buffer budget for the data pipeline. This is generally applicable, and ParaGAN enables this feature by default.  

\begin{figure}[t]
  \centering
  \includegraphics[width=\linewidth]{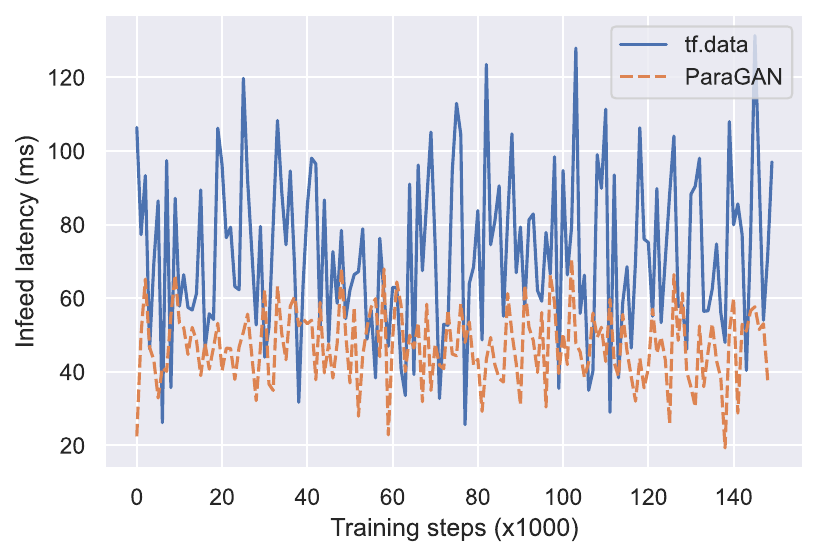}
  \caption{Data pipeline latency.}
  \label{fig:pp_latency}
\end{figure}

We compare the performance of our congestion-aware pipeline with TensorFlow's native data pipeline (tf.data). To ensure the results are comparable, they are run at the same time on the same type of machine pointing to the same dataset storage node, and latency is measured at the time taken to extract a batch of data. As shown in Fig.~\ref{fig:pp_latency}, our pipeline tuner has a lower variance in latency.

\textbf{Low-precision (with bf16)} training provides additional 14-17\% performance gain and reduces TPU memory usage by 24\%. Bfloat16 format saves activation values in lower bits, which makes it faster to load from memory and communicate with other workers. We have observed while converting activation to lower precision does not harm convergence, numerical instability may happen when converting gradient and weight into bfloat16. In ParaGAN, the users can control the precision for each layer.

\textbf{Layout transformation} provides 4\% additional improvement by increasing the accelerator utilization and reducing kernel launch overhead. This is achieved using a profiling-guided approach, as we observe that many smaller tensors are fed into the same convolution kernel. Considering that it optimizes the data layout, it is possible to integrate it into the XLA as an HLO pass.

\subsection{Generating High-Resolution Images}
To our knowledge, we are the first to successfully train BigGAN at $1024\times1024$ resolution, which is 4 times higher than the result shown in the original BigGAN. Training at high resolution is particularly challenging because the generator will need to use more channels and deconvolutional layers to generate more details. It is therefore more sensitive to hyperparameters and initialization. Different from ProgressGAN ~\cite{karras2017progressive} where they use progressive growing to train low-resolution images first before increasing the resolution, we directly train it on $1024\times1024$ resolution, which is more challenging, and it requires the numerical optimization techniques we discussed. 

\begin{figure}[t]
\includegraphics[width=\linewidth]{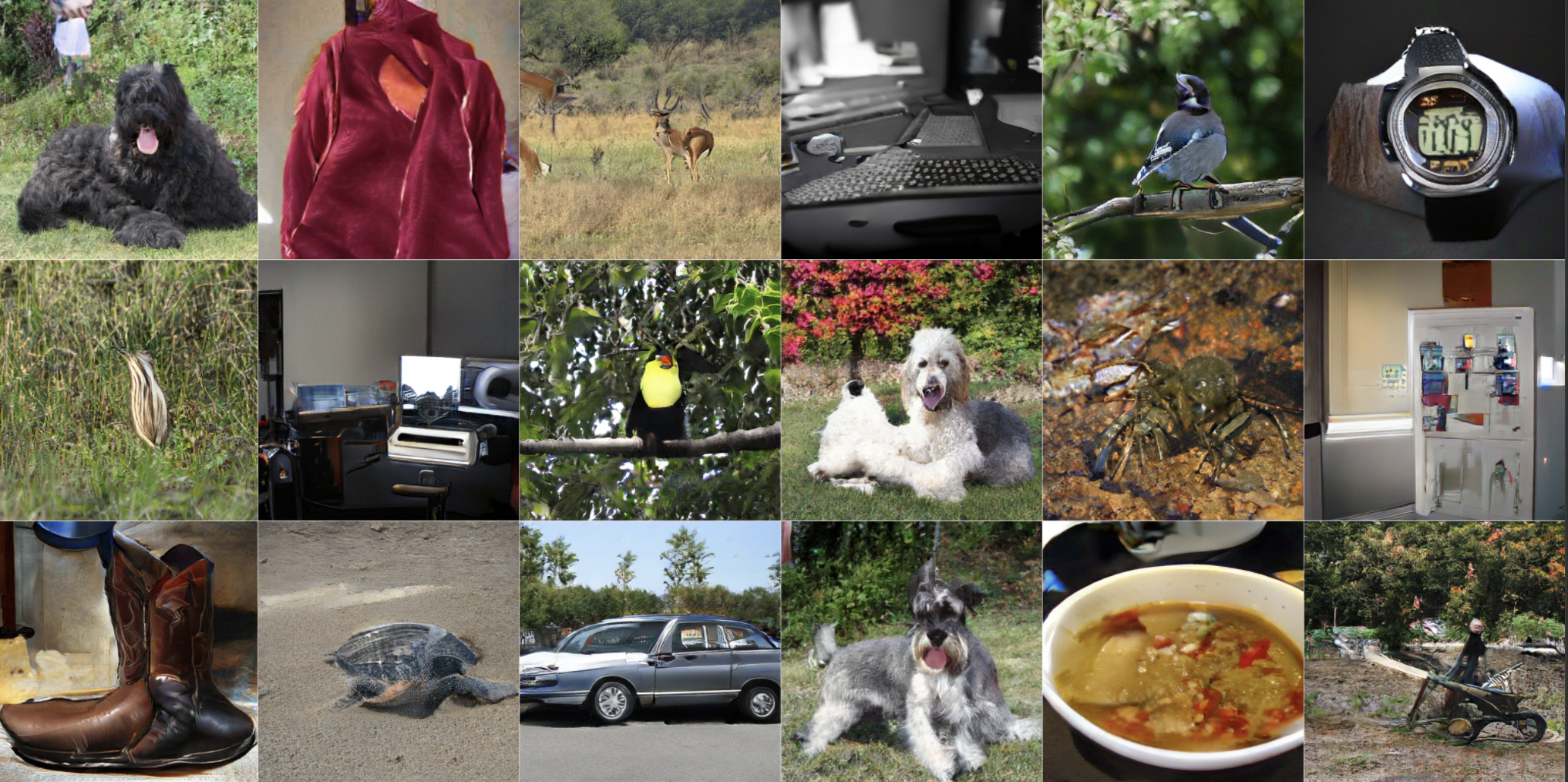}
\centering
\caption{Output of BigGAN at 1024x1024 resolution. Best viewed in color.}
\label{fig:gan_output}
\end{figure}

The generated results (trained on ImageNet dataset) achieved an Inception Score (IS) ~\cite{salimans2016improved} of 239.3 and Fréchet Inception Distance (FID) of 13.6. They are presented in Fig. \ref{fig:gan_output} for visual evaluation.

\subsection{Convergence of Async. Update Scheme}

\begin{figure}
     \centering
     \begin{subfigure}[b]{\linewidth}
         \centering
         \includegraphics[width=\linewidth]{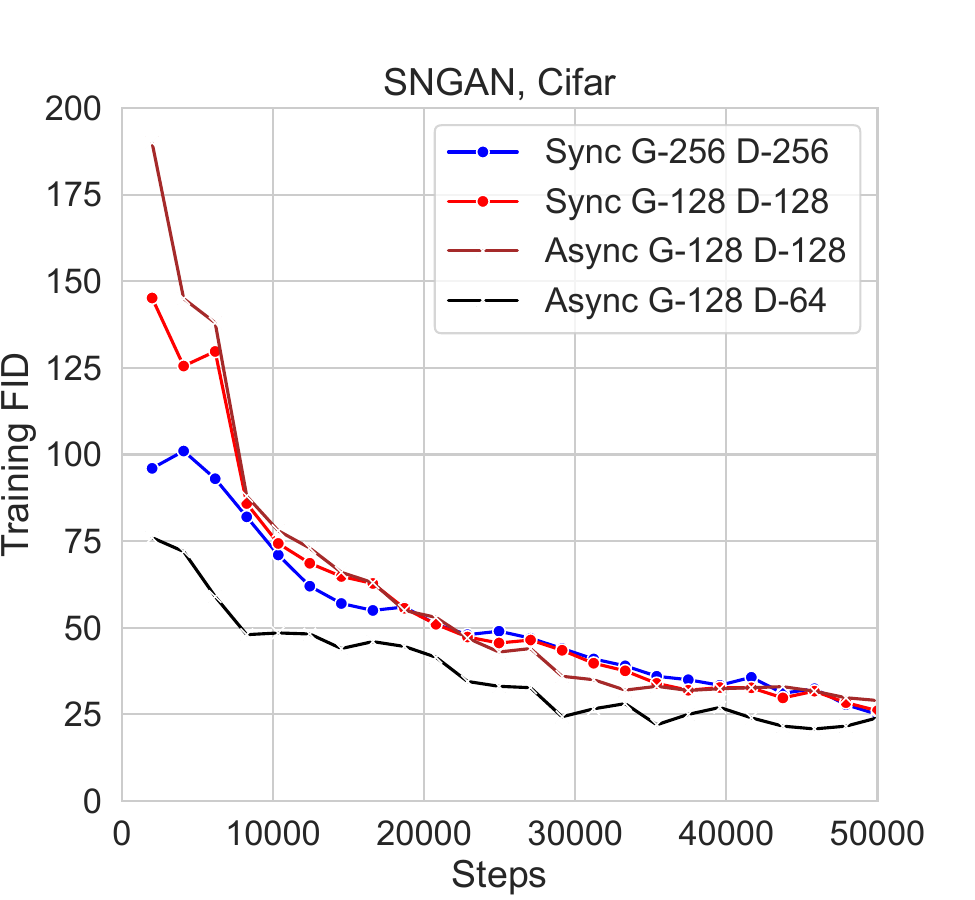}
         \caption{Cifar.}
         \label{fig:cifar}
     \end{subfigure}
     \hfill
     \begin{subfigure}[b]{\linewidth}
         \centering
         \includegraphics[width=\linewidth]{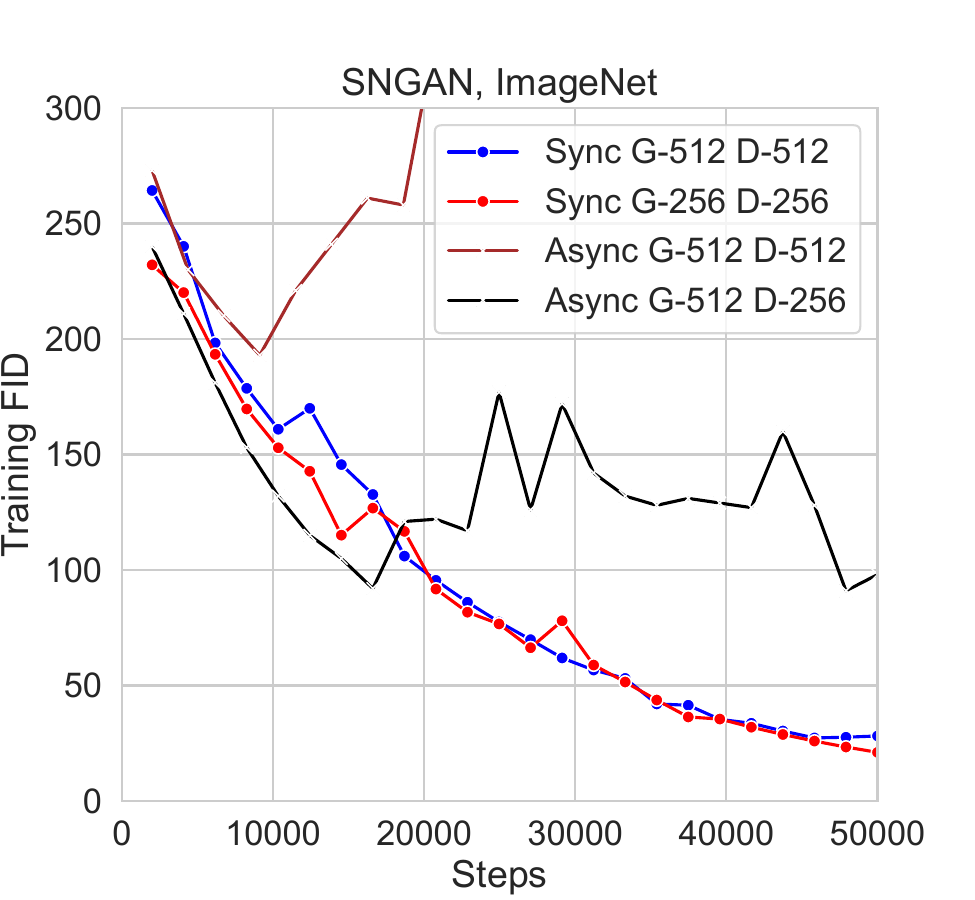}
         \caption{ImageNet.}
         \label{fig:img}
     \end{subfigure}
    \caption{The FID score of images produced using the Async Update scheme (lower is better). The number in the legend represents batch size.}
    \label{fig:fid}
\end{figure}

We study the convergence behavior of the asynchronous update scheme on SNGAN and present in Fig.~\ref{fig:fid}. We observe that the asynchronous update scheme can accelerate convergence on smaller datasets, and the benefit is more obvious in the early stage of training. However, the asynchronous update scheme (Async G-512 D-256) struggles to converge on high-resolution image generation tasks despite reaching lower FID quicker than the synchronous versions before 16K steps. Throughout repeated experiments and other models, we observe similar results. We conjecture that the asynchronous update scheme may be used to accelerate the earlier stage of training before switching to the synchronous update scheme for better convergence. 

We recognize that theoretical convergence analysis in the context of GAN training is an active area of research~\cite{heusel2017gans, kodali2017convergence, Barnett2018ConvergenceGANs,mescheder2018training, Niu2011HOGWILD:Descent}, and is particularly challenging. As GANs operate on a two-party optimization problem, their convergence is inherently difficult to guarantee, especially when using stochastic gradient descent. Furthermore, introducing asynchrony, while beneficial in terms of performance, adds another layer of complexity in analyzing convergence and stability. While we have relied on empirical results to demonstrate the effectiveness of these techniques, we leave it as future work to formalize a theoretical framework for asynchronous GAN training. 

\section{Related Work}\label{sec:related_work}

\subsection{Generative Adversarial Network}

Several methods for GANs are proposed to improve the performance and training stability of GAN. SNGAN ~\cite{miyato2018spectral} uses spectral normalization to stabilize the training of the discriminator and achieves better inception scores relative to previous studies. BigGAN ~\cite{Brock2019LargeSynthesis} scales up the batch size and the model size for training to generate high-resolution and high-quality images. StyleGAN ~\cite{karras2019stylebased} enhances the generator model to produce photorealistic faces with high resolution meanwhile being able to control the style of the generated image through varying the style vectors and noise. However, the above works were less focused on the training acceleration aspect.

\subsubsection{Mode Collapse}

Mode collapse is a common type of GAN failure where the generator fails to produce a good variety of samples that can fool the discriminator. When it happens, the generator sticks with a set of similar-looking output patterns without progressing, and the discriminator can almost always classify the generated samples. Essentially, the generator's optimization process reaches a saddle point, and rotates through the small set of output. Several approaches have been proposed to try to solve the problem. VEEGAN ~\cite{srivastava2017veegan} features a reconstruction network reversing the generator by mapping samples to random noise. Unrolled GAN ~\cite{metz2017unrolled} uses a loss function for the generator concerning an unrolled optimization of the discriminator. This way, the generator training can be adjusted by both current outputs and optimal outputs of the discriminator. Still, mode collapse remains challenging to identify and mitigate.

\subsection{Distributed GAN Training}
Existing works on distributed training for GAN are relatively limited with a focus on the privacy-preserving perspective. Since GAN consists of two sub-networks, a common approach is to let multiple discriminators serve one centralized generator using parameter server~\cite{Hardy2019MD-GAN:Datasets, Qu2020LearnDiscriminators, chang2020synthetic}. 
The central generator sends synthesized output images to the distributed discriminators, and the discriminators update the generator with their predictions. FeGAN~\cite{guerraoui2020fegan} instead deploys a complete GAN on each device to address the data skewness and mode collapse issue.
AI-GAN~\cite{Jin2020AI-GAN:Removal} tackles the image-deraining problem by using a two-branch network that learns a disentangled representation for the rain and background, and jointly optimizes the two branches via mutual adversarial optimization. 
~\cite{toutouh2020parallel} adopts the federated learning setting on the medical image generation problem on low-resolution images. 
\cite{yang2019highly} proposes a PDE-informed GAN architecture for subsurface flow characterization problem, achieving 93.5\% scaling efficiency on 27500 GPUs with communication and load balancing optimizations. 
\cite{liu2020decentralized} presents a decentralized training algorithm named DPOSG that caters to the non-convex non-concave min-max optimization process. 

To the best of our knowledge, ParaGAN is the first system that efficiently scales GAN on high-resolution image generation tasks on a large-scale cluster.

\section{Conclusion}\label{sec:conclude}

ParaGAN is a large-scale distributed GAN training framework that supports high-resolution image generation with near-linear scalability. ParaGAN is optimized with a dynamic data pipeline, mixed-precision training, and layout transformation. We show that it is possible to train the generator and discriminator independently using an asynchronous update scheme and asymmetric optimization policy. ParaGAN scales almost optimally to 1024 accelerators, and it can greatly reduce the time to train a GAN model from weeks to hours. We believe ParaGAN can advance GAN research by accelerating the training process. 
\section{Acknowledgement}
This research is supported by the Ministry of Education AcRF Tier 1 grant (No. T1 251RES2104) and Tier 2 grant (No. MOE-T2EP20222-0016). This is also supported by the Cloud TPUs from Google's TPU Research Cloud (TRC).

\bibliographystyle{socc24-final-template/ACM-Reference-Format}
\bibliography{socc24-final-template/references.bib}

\end{document}